\documentclass[twocolumn,twoside,slac_two]{revtex4}
\usepackage{graphicx}
\usepackage{fancyhdr}
\newcommand{\kms}{km~s$^{-1}$}

\newcommand{\etal}{{\it et al.}}

\newcommand{\be}{\begin{equation}}
\newcommand{\ee}{\end{equation}}
\newcommand{\arcdeg}{$^\circ$}
\newcommand{\kmsMpc}{km~s$^{-1}$ Mpc$^{-1}$}

\begin{document}

\title{Cosmology in the Very Local Universe - Why Flow Models Matter}
\author{Karen L. Masters}
\affiliation{Department of Astronomy, Cornell University, Ithaca, NY, 14853, USA}

\begin{abstract}
While much of the focus of observational cosmology is on the high redshift universe it is important not to neglect the very local universe as a source of cosmological information. The inner profiles and number counts of low mass halos have provided the biggest stumbling blocks so far for $\Lambda$CDM. These small structures can only be seen nearby. In the very local universe ($cz < 3000$ \kms) the component of a galaxy's redshift which is due to motions under gravity can be comparable to (or even larger than) its cosmological redshift. The distance to a galaxy as inferred from its redshift can differ by more than a factor of two from its actual distance. Given that the mass and intrinsic size scale of a galaxy (among many other physical parameters) depend strongly on distance, and that these peculiar motions are coherent over large regions of the sky, serious biases can occur. While it is important to have accurate distances to local galaxies, it is not feasible to measure a primary distance for every nearby galaxy. Instead, a velocity field model can be used to provide a first order correction to the redshift distance. We report on a new sample of Tully-Fisher distances (the SFI++) which is being used in combination with publicly available primary distances to model galaxy flows in and around the Local Supercluster. This sample has $\sim$10 times as many tracers as were used for the current best model. Initially a parametric model including infall onto multiple attractors will be used. Such models (which assume spherical symmetry for the attractors) are not realistic representations of the true velocity field, but provide useful first order corrections. We explore the typical errors in distances from such a flow model and how they vary in different regions of the local supercluster by fitting the same parametric model to mock catalogs derived from a constrained simulation of the volume. Non-parametric reconstructions of the velocity field will follow. 
\end{abstract}

\maketitle

\thispagestyle{fancy}

\section{Introduction}
Cosmology has entered a new era. Recently large redshift surveys have been used in combination with observations of the cosmic microwave background to provide such strong constraints on the cosmological parameters that many astronomers now consider cosmology `solved' (eg. \cite{T04}), giving the concordance model of $\Lambda$CDM in which the universe consists of 4\% baryons, 26\% dark matter and 70\% dark energy. Much of this information has come from studies of the very high redshift universe - for example the WMAP observations of the fluctuations in the cosmic microwave background \cite{WMAP}, or the studies of high redshift Type Ia supernova that gave the first evidence for dark energy \cite{SNIa2}, \cite{SNIa1}. 

 On the large scales $\Lambda$CDM has stood up extremely well to observational tests. Its biggest challenges have so far come from much smaller scales; for example in measurements of the inner density profiles and number counts of low mass galaxies. Such observations can only be done in the local universe, and therefore are impacted by the uncertainty inherent in estimating distances in this region where Hubble's law cannot be used. There are many thousands of galaxies with measured redshifts less than a few thousand \kms, so it is obviously not feasible to measure redshift independent distances to all of them. Flow models are therefore the only was to provide unbiased distances to these galaxies.

 In this paper we will discuss the impact the local peculiar velocity field has on cosmology, and the use of simple flow models to estimate galaxy distances in the local universe. In Section 2 we will describe the local velocity field and the pattern of deviations from Hubble's Law. In Section 3 we will discuss an example of the impact of ignoring peculiar velocities, specifically on the inferred number counts of low mass galaxies. In Section 4 we will introduce the SFI++, a new sample designed to study the local velocity field. In Section 5 we provide a preliminary report on a test of distances estimated using simple infall models, which is done by fitting such a model to a mock catalog generated from a constrained simulation of the volume. Section 6 provides a summary of the paper.

\section{The Local Velocity Field}
\begin{figure*}
\includegraphics[width=100mm]{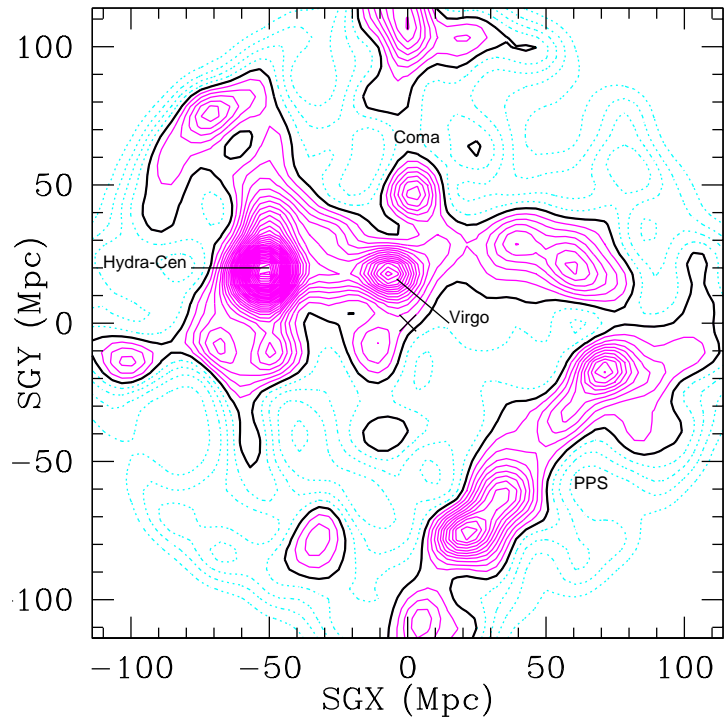}
\caption{A slice of the density field in the Supergalactic plane from the constrained simulations described in \cite{Mathis}.  The density field has been smoothed with a Gaussian of radius 4 Mpc. Contours are spaced at intervals of $\delta = (\rho - \rho_0)/\rho_0 = 0.2$. The mean density of the universe is shown by the thick solid line, while the magenta solid lines show overdensities, blue dotted lines underdensities. }
\label{slice}
\end{figure*}
 The uneven distribution of mass in the universe gives galaxies deviations from Hubble flow (or peculiar velocities) of the order 200-600 \kms. In the local universe cosmological redshifts are small, so these peculiar velocities can dominate the observed recessional velocity of a galaxy. Gravitational instability theory in the linear regime in an expanding universe shows that the peculiar velocity at any given point in space is directly proportional to the gravitational potential. 
\be
{\bf v}({\bf x}) = \frac{2}{3 H_0 \Omega_{\rm M}} f  {\bf g}({\bf x})
\ee
where the proportionality constant depends on cosmological parameters, and the factor $f \sim \Omega_{\rm M}^{0.6}$ which describes the rate of growth of structures. N-body simulations have shown that these motions are dominated by the infall of galaxies along filaments towards the clusters which form at the intersections of filaments. 

 Our nearest large cluster is the Virgo Cluster at a distance of $\sim$16 Mpc away, and within the Local Supercluster (of size $\sim$30 Mpc) galaxies lie preferentially in a plane, known as the supergalactic plane\cite{V58}. At our location the infall velocity onto Virgo is $v_{\rm inf} \sim$200 \kms, but this infall cannot account for all of the $v=368\pm$1.8 \kms ~peculiar velocity of the Local Group in the CMB frame (as measured by COBE and now WMAP \cite{WMAP}), largely because it points in a direction $\sim$50\arcdeg ~away from the direction to the Virgo cluster. This and other observations of the motions of galaxies in the local universe have revealed a flow in a the direction towards the Hydra-Centaurus cluster and Shapley Supercluster in the background (or the ``Great Attractor" region), as first suggested by \cite{LB88}. 

The distribution of mass in a slice through the supergalactic plane from the constrained simulations described in \cite{Mathis} is shown in Fig \ref{slice}, with the major nearby concentrations of mass indicated. This slice goes out in distance to $cz = 8000$ \kms ~and has been smoothed with a Gaussian of radius 4 Mpc ($H_0 = 70$ \kmsMpc ~in the simulation).

\section{The Missing Satellite Problem}
 One of the largest discrepancies between observations and $\Lambda$CDM revolves
around the ``missing satellite'' problem.
Numerical simulations in $\Lambda$CDM predict a value for the logarithmic slope of the faint end of the halo mass function, $\alpha \sim -1.8$, while most recent determinations of the optical 
luminosity function (LF) in the local volume 
yield values of $\alpha$ that are significantly flatter.
Complementary to the results on the optical LF, several determinations of the
local HI mass function (HIMF) have likewise yielded relatively flat faint-end slopes.
The over-prediction of the number of low mass objects relative to those actually observed 
is considered one of the last remaining stumbling blocks for $\Lambda$CDM, and points 
towards a new understanding of the baryon physics of galaxy formation. It is obviously 
desirable that all of the uncertainties in the observed luminosity and mass functions 
be well understood.

\begin{figure*}
\includegraphics[width=95mm]{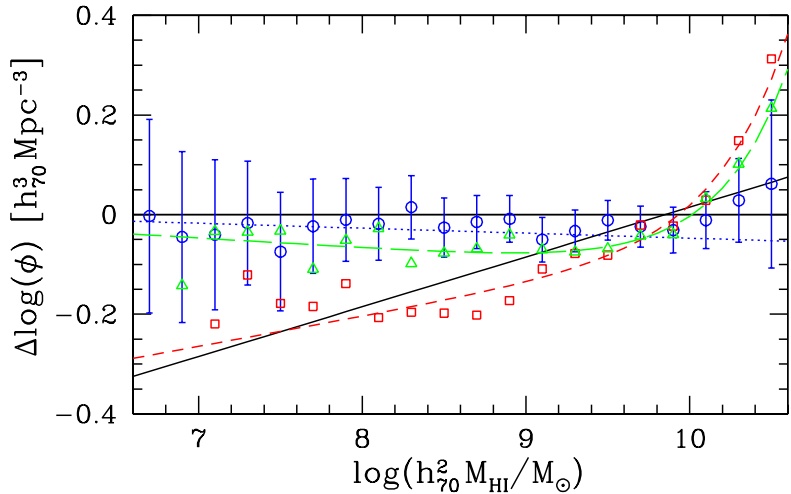}
\caption{Difference between input and reconstructed HIMFs for a mock HIPASS BGC survey \cite{Z03}, (shown are the average 
of 10 simulations).  The HIMF derived when ``true'' distances are used (blue circular points and dotted line), along with the HIMF constructed when multi-attractor model distances are used (green triangular points and long-dashed line) reconstructs the input HIMF within the errors (here typical Poisson counting errors are shown only on the ``true'' distance points). When pure Hubble flow is assumed (red square points and short dashed line), the low mass end of the HIMF is underestimated. The solid line shows the HIMF derived in \cite{Z03}, which differs from the input by $\Delta \alpha = 0.1$. }
\label{himf}
\end{figure*}

To derive a luminosity (or mass) function from observational data, the sample selection effects 
must be considered carefully. In most surveys, the lowest luminosity 
(or mass) objects are only visible nearby, and their contribution must be 
weighted accordingly. 
Fractional distance errors can be large for relatively nearby
objects, and can thus have a very strong impact on determinations
of the faint end of luminosity (or mass) functions. In particular, the derivations of HIMFs have to rely on relatively shallow, wide area surveys to detect low HI mass objects over a sufficient volume, and are therefore most susceptible 
to uncertainties associated with distance errors in the nearby universe.

 In \cite{M04} we described simulations of the impact of neglecting peculiar velocities on the derivation of HIMFs from recent HI surveys. We showed that the HIMF derived from the HIPASS Brightest Galaxy Catalog \cite{Z03} underestimated the slope of the low mass end of the HIMF because of their use of Hubble's law to estimate distances to nearby galaxies (see Fig \ref{himf}). HIPASS is a HI survey of the southern sky. Nearby galaxies in the south will systematically have their distances (and therefore masses for a given HI flux) overestimated if peculiar velocities are neglected, largely because of their motion towards the ``Great Attractor'' region at (RA, DEC) $\sim$ (13 hr, -35\arcdeg). This systematic effect results in an underestimate of the number counts of low mass galaxies as inferred from the HIMF derived from this data. 

 The size of the bias is not nearly large enough to explain the missing satellite problem, but it serves to illustrate the point that without a good understanding of the local peculiar velocity field mistakes can be made which influence our understanding of cosmology as a whole.

\section{The SFI++} 

\begin{figure*}[t]
\centering
\includegraphics[width=125mm]{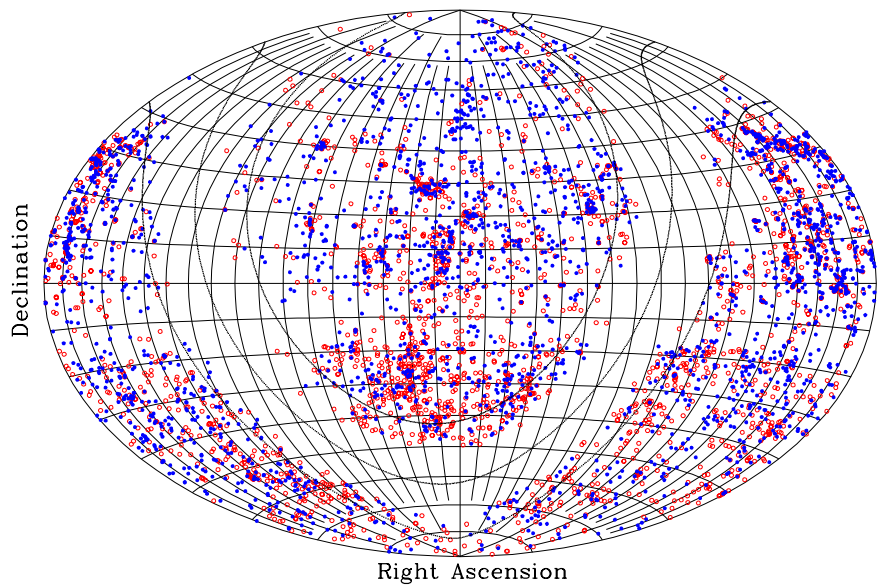}
\caption{The sky distribution of galaxies in the SFI++ sample centered at RA=12 hours. Galaxies are shown as blue filled circles if they have a negative peculiar velocity, red open circles if they have a positive peculiar velocity. The lines show the location of the Galactic plane. }
\label{SFI1}
\end{figure*}

 The SFI++ is a sample of spiral galaxies with I-band photometry and a combination of HI and H$\alpha$ spectroscopy suitable for use in the Tully-Fisher (TF) relation (a relation between the absolute magnitude and rotation velocity of a spiral galaxy). This sample builds on the all-sky SFI (Spiral Field I-band) sample (\cite{G97a},\cite{G97b},\cite{H99a},\cite{H99b}) and contains much new I-band photometry, HI and H$\alpha$ spectroscopy in areas of the sky north of -20\arcdeg. Data from the SFI has also been reprocessed to provide a uniform sample. The SFI++ will provide Tully-Fisher distances for $\sim 5000$ galaxies out to $cz = $ 10 000 \kms. It is a diameter limited catalog, with diameter limits which vary with the redshift of the galaxy in a similar way to the SFI, except that the minimum diameter limit for SFI++ is smaller. Figure \ref{SFI1} shows the preliminary sky distribution of the galaxies in this sample. 

 A multiattractor model is being fit to a subset of these Tully-Fisher distances (for galaxies within $cz = $4000 \kms) combined with a sample of $\sim$ 500 publicly available primary distances collected from the literature. Primary distances provide lower errors for individual measurements, but larger numbers of velocity field tracers can be obtained using secondary indicators like TF. This combination will provide the best currently available sample to study the local velocity field and represents a significant increase in the number of velocity field tracers from what has previously been used. The multiattractor model should not to be considered a realistic representation of the local density and velocity fields, but will provide first order corrections to galaxy distances estimated from their redshifts. Non-parametric modeling of the velocity field (e.g. POTENT \cite{Dekel}) which can provide a more realistic picture will follow.

\section{Testing Multiattractor Models on Constrained Simulations}
 Numerical simulations in the $\Lambda$CDM paradigm and redshift surveys have given us a consistent picture of the filamentary large scale structure of the universe in which the dominant motions of galaxies are that of infall onto clusters
along filaments. In light of this picture the traditional infall models used to
study the velocity field of the Local Supercluster going back to \cite{S80}, and most recently used by \cite{TB00} and \cite{WB01} are at best over simplifications and at worst simply wrong. The assumption of spherical symmetry for attractors is clearly an over simplification, even for a single cluster like Virgo which is made up of many sub-groups, let alone for a filamentary or pancake like supercluster.
                                                                                
 However, multiattractor models have many nice features which leads to their continuing use. They are simple to apply, and the non-linear approximation of \cite{Y85} (which is only valid for spherical masses) has been shown to work reasonably well to overdensites of $\delta \sim 30$ \cite{GM93}. For a galaxy which has only a position on the sky and a redshift, a multiattractor model will provide a unique distance over much (if not all) of its volume. A multiattractor model can in principle have any number of spherical attractors which could be combined to give more complicated mass distributions, and other components, such as quadrupole moments in the velocity field (perhaps from the tidal influence of distant masses) are also easy to add. A multiattractor model can also give you an idea of the dominant sources of attraction in the local volume, taking into account a combination of the total mass and distance to the attractors.

 Here we test the validity of a simple multiattractor model as applied to a mock distance catalog generated from the Virgo Consortium's publicly available constrained simulations of the local universe \citep{Mathis}. We test distance predicted from the best fit multiattractor model against the real distances in the constrained simulation to see how well such a model can be used to predict distances for galaxies in the local universe.

\subsection{Mock Catalog of Distances}

\begin{figure*}
\includegraphics[width=85mm]{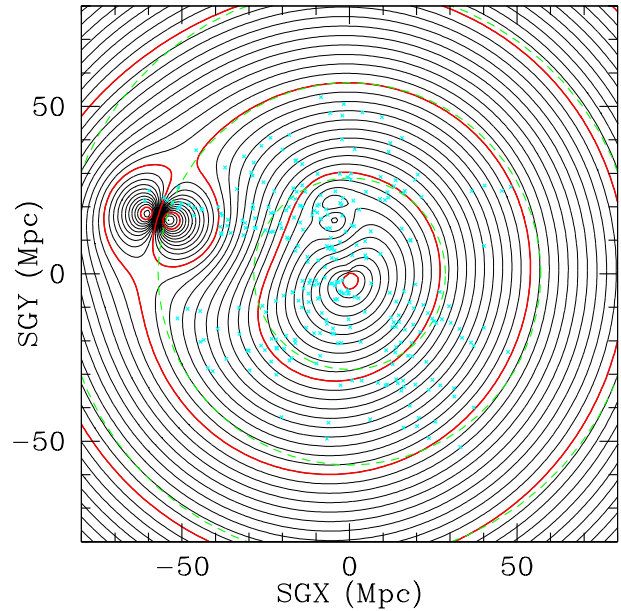}
\caption{Contours of constant heliocentric velocity in the supergalactic plane for the best fit multiattractor model of a mock sample of distance indicators. The location of the mock galaxies used in the fit are also shown. Thin black contours are places at 200 \kms ~ intervals with the thick red contours every 2000 \kms. The concentric green dashed circles show pure Hubble flow with $H_0=70$ \kmsMpc. }
\label{model}
\end{figure*}

 The mock catalogs used here are based on the constrained simulations described in \cite{Mathis}. These $\Lambda$CDM simulations are designed to mimic the local density and velocity field. Semi-analytic galaxy formation \cite{K99} is grafted onto the N-body simulation to create galaxy catalogs. We generate simple mock catalogs of distance indicators from these publicly available catalogs by imposing a magnitude limit of $m_I = 12$ and selecting a random sample of $\sim$ 500 such galaxies with $cz < 4000$\kms. The sample is uniform over the sky except that no galaxies are seen within 10\arcdeg ~of the Galactic plane. Gaussian scatter of width 0.2 mag is added to the distance modulus to simulate 10\% distance errors.

\subsection{The Best Fit Multiattractor Model}

 We fit a multiattractor model to the mock distance catalog described above. This model has a very similar functional form to that described in \cite{TB00}, except that we use NFW profiles for the attractors. The best fit model includes infall onto both Virgo, and the ``Great Attractor'' region and a quadrupole component to the velocity field which might account for the tidal influence of external masses. 
Fig \ref{model} shows contours of constant heliocentric velocity in the supergalactic plane from this model, along with the locations of the mock galaxies it was fit to.

\subsection{Predicting Distances from a Flow Model}

 A flow model gives you the peculiar velocity of a galaxy at a given location. In order to use it to give distances for galaxies with observed redshifts one must invert it, and therefore solve an equation of the form $D \nolinebreak = \nolinebreak f(D, v_{\rm obs}, {\rm RA}, {\rm DEC})$. We do this using Numerical Recipes root finding algorithms. Galaxies which are close to the location of attractors are assumed to be in the attractor core. Galaxies in the triple values regions (where three distances along a line of sight have the same observed recessional velocity) are assigned to the central location, and the distance error is inflated to bracket all three possible distances. 

\subsection{Flow Model Distance Errors}

\begin{figure}
\includegraphics[width=65mm]{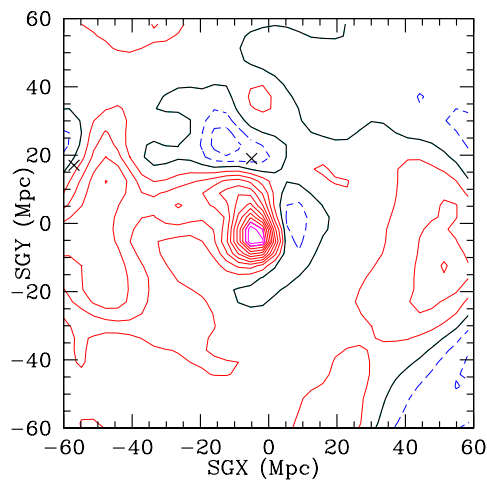}
\caption{Contours of the mean percent error in distances estimated using the multiattractor model for galaxies within 5 Mpc of the supergalactic plane. Errors are smoothed on a scale of 4 Mpc. The contour spacing shows a 5\% error with red solid contours being overestimated distances, blue dahsed contours underestimated.  }
\label{flowresid}
\end{figure}

\begin{figure}
\includegraphics[width=65mm]{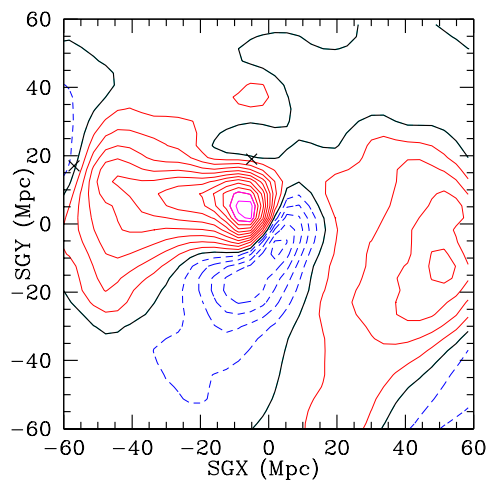}
\caption{As in Fig \ref{flowresid} except for distances estimated assuming Hubble's Law. }
\label{Hubbleresid}
\end{figure}

 The use of a this flow model gives distance estimates to galaxies to less than 5\% accuracy over much of the local volume. This is illustrated in Fig. \ref{flowresid} which shows contours of the average percent error for distances estimated using the flow model. Fig. \ref{Hubbleresid} shows the same thing, but this time for distances estimated using Hubble's Law. Note that near the origin where the absolute distance is small there will always be an area where distances are overestimated by large percentages (it is not possible to underestimate distances to galaxies which are very close by). The flow model does not completeley correct for the area of systematically overpredicted distances in the direction towards the ``Great Attractor'' region at (SGX, SGY) $\sim$ (-50, 20) Mpc, nor can it completely correct for the velocity field in the direction towards Virgo at  (SGX, SGY) $\sim$ (-3, 20) Mpc. 

\section{Summary}

 We have argued that the local universe is an important source for cosmological information, especially in light of the fact that the two largest challenges still faced by $\Lambda$CDM come from observations which are only possible in the local universe. 

 Within a redshift of $cz \sim$ 4000\kms ~Hubble's law cannot be used to predict distances. We show, using the example of the derivation of the HI mass function for galaxies in the southern sky, that if Hubble's law is used to derive distance serious biases can occur. This work is presented in detail in \cite{M04}.

 We present the SFI++, and all-sky sample of $\sim$5000 spiral galaxies for which Tully-Fisher distances will be available. This sample builds on the SFI sample of the 1990s (\cite{G97a},\cite{G97b},\cite{H99a},\cite{H99b}) and provides the best currently available sample for studies of the local peculiar velocity field. Initially a multiattractor model will be fit to this data, more realistic modeling will follow.

 We provide a preliminary report of work testing the distances predicted the using commonly applied, but simplistic multiattractor infall models. This work makes use of the publicly available Virgo Consortium constrained simulations \cite{Mathis}, and fits a multiattractor model to a mock sample of 500 galaxies with distances measured to 10\% derived from this simulation. The best fit model includes infall onto both the Virgo cluster, and Hydra-Centaurus (`Great Attractor') region, as well as a small quadrupolar component to the velocity field. Preliminary results suggest that this multiattractor model can be used to estimate distances to better than 5\% on average over much of the local volume, but care must be taken in the directions towards both attractors.

\bigskip 
\begin{acknowledgments}
The author would like to thank her adviser, Martha Haynes for advice on this conference presentation, and all members of the Cornell EGG who have been involved in assembling data for the SFI++ sample. This work has been partially supported by NSF grants AST-0307396 and AST-0307661.
\end{acknowledgments}

\end{document}